# The collision theory reaction rate coefficient for power-law distributions

Yin Cangtao and Du Jiulin

*Department of Physics, School of Science, Tianjin University, Tianjin 300072, China*

**Abstract:** The collision theory for power-law distributions and a generalized collision theory rate coefficient is studied when the reactions take place in nonequilibrium systems with power-law distributions. We obtain the power-law rate coefficient and by numerical analyses we show a very strong dependence of the rate coefficient on the power-law parameter. We find that the power-law collision theory can successfully overcome the two difficulties of Lindemann–Christiansen mechanism. We take three reactions as examples to calculate the pre-exponential factor and yield the values that can be exactly in agreement with those measured in the experimental studies.

**1. Introduction**

Calculations of reaction rate coefficients are of an inter-discipline of nonlinear science, and are very important for us to study and understand many basic problems appearing in many different physical, chemical, biological and technical processes. There have been various reaction rate theories that have been developed to calculate reaction rate coefficients, among which the collision theory is an old and foundational one [1]. Other reaction rate theories, such as transition state theory, Kramers rate theory, and unimolecular rate theory, all borrow the idea from the collision theory [1-3]. More important, analysis of the collision phenomena plays a central role in almost all investigations of structures of matters on microscopic scale. For the collisions between two molecules A and B, all molecules are assumed to comply with a statistical distribution of thermodynamic equilibrium, and thus Boltzmann-Gibbs (BG) statistics naturally becomes the statistical base of the collision theory. In this way, conventionally the collision theory reaction rate coefficient is given in the form with an exponential law [1] by

$$k_{col} = \pi d_{AB}^2 \sqrt{\frac{8k_B T}{\pi \mu}} \exp(-\beta \varepsilon_c), \qquad (1)$$

where $d_{AB}$ is the distance between the centers of molecules A and B, $k_B$ is Boltzmann constant, $T$ is temperature, $\mu$ is a reduced mass defined with the masses of A and B, $m_A$ and $m_B$, by $\mu = m_A m_B / (m_A + m_B)$, $\varepsilon_c$ is the critical energy of a molecule at which the reaction happens, and $\beta = (k_B T)^{-1}$ is Lagrangian multiplier.

However, as we know, chemical reactions are generally far away from equilibrium, the statistical property may not follow BG statistics and therefore does not have an exponential-law distribution. A lot of theoretical and experimental works have shown that power-law distributions are quite common in the physical, chemical, biological and technical processes taking place in complex systems (see [4] and the references therein). In a reacting system, particles escape over the barrier would result in a perturbation about the Maxwell-Boltzmann distribution in the well [2]. Here we can introduce the power-law distribution in nonextensive statistical mechanics [5]. The power-law $\nu$-distribution can be written by

$$P(\varepsilon) \sim [1-(\nu-1)\beta\varepsilon]^{\frac{1}{\nu-1}}, \qquad (2)$$

if the energy $\varepsilon$ is small. Or, we can write $P(\varepsilon) \sim \varepsilon^{-\alpha}$ if the energy $\varepsilon$ is large [4]. The power-law $\nu$-distribution represents the statistical property of a complex system being at a nonequilibrium stationary-state [6, 7]. Eq.(2) can be reduced a BG distribution if the $\nu$-parameter is set $\nu \to 1$, where the parameter $\nu \neq 1$ measures a distance away from equilibrium [4]. The power-law distributions in complex systems have been noted prevalently in the processes such as single-molecule conformational dynamics [8, 9], , chemical reactions [10-12], gene expressions [13], cell reproductions [14], complex cellular networks [15], and small organic molecules [16] etc. In these processes, the reaction rate coefficients may be energy-dependent (and/or time-dependent [17, 18]) with power-law forms [19, 20], which are beyond the reaction rate formulae in the collision theory govern conventionally by the BG exponential laws. In these cases, the reaction rate formulae become invalid and so need to be modified. Most recently, the transition state theory was generalized to the nonequilibrium systems with power-law distributions [19], and the power-law reaction rate coefficient was studied for an elementary bimolecular reaction [21]. In addition, the nonextensive survival probability and the associated Kramers rate were studied by using nonextensive formalism [22], the mean first passage time for power-law distributions [23] and the escape rate for power-law distributions in both overdamped systems and low-to-intermediate damping [24,25] were also studied. As we can imagine, this is a complicated and exciting field in exploring the understanding of nonequilibrium reaction rate theory. The purpose of this work is to generalize the collision theory reaction rate formula to a nonequilibrium system with power-law distributions.

The paper is organized as follows. In Sec.2, we study the collision theory for the power-law distribution and derive the rate power-law coefficient formula. In Sec.3, we make numerical analyses of the dependence of the new rate coefficient on the quantities such as $\nu$-parameter, temperature and critical energy etc. In Sec.4, we apply the new theory to the Lindemann–Christiansen mechanism [26]. In Sec.5, we take three examples of chemical reactions to calculate the pre-exponential factor and to compare with the experiment studies. Finally, in Sec.6 we give conclusions and discussions.

**2. The power-law collision theory**

As a first step of the generalization of the collision theory rate formulae to the complex systems with power-law distributions, we follow the standard line in textbooks to derive the reaction rate formula in the power-law collision theory. Let us consider the simple collision theory. Because it is not actually satisfactory as a theoretical hypothesis for polyatomic systems, so we will restrict this type of calculation to a simple system.

To calculate the collision number per unit time in the system, we need a molecule model. The simplest approach involves a system of two gases, A and B, whose molecules behave as hard spheres characterized by the impenetrable radii $r_A$ and $r_B$. The collision between A and B occurs when their centers approach within a distance $d_{AB}$, such that $d_{AB} = r_A + r_B$. If we assume that the molecules of B are fixed and those of A move with an average velocity $\bar{u}_A$, each molecule A sweeps a volume $\pi d_{AB}^2 \bar{u}_A$ per unit time which contains stationary molecules of B. The area $\sigma_p = \pi d_{AB}^2$, is known as the collision cross section. If there are $N_B/V$ molecules of type B per unit volume, the number of collisions of a molecule of type A with the stationary molecules B will be $z_{AB} = \pi d_{AB}^2 \bar{u}_A N_B / V$. If the total number of molecules of A per unit volume is $N_A/V$, then the total number of collisions of A with B per unit volume (collision density) is [1] given by

$$Z_{AB} = \pi d_{AB}^2 \bar{u}_A \frac{N_A N_B}{V^2}, \tag{3}$$

where, as indicated above, we have assumed that the molecules of B are stationary to obtain the expression. In practice, for each pair of molecules A and B involved in a collisional trajectory, we can define a relative velocity $u_{AB}$, which is related to their velocities $u_A$ and $u_B$ by $u_{AB}^2 = u_A^2 + u_B^2 - 2u_A u_A \cos\theta$. The value of $\cos\theta$ can vary between -1 and 1. As all values of $\theta$ between 0 and $2\pi$ are equally probable, the positive and negative values of $\cos\theta$ will cancel out for the square of $u_{AB}$, and the mean value will be zero, so one obtains $\bar{u}_{AB}^2 = \bar{u}_A^2 + \bar{u}_B^2$ [1]. We would have $u_A = u_{AB}$ if we assumed the molecule B is stationary.

If the reactant molecules were assumed to comply with a statistical distribution at a thermodynamic equilibrium state, Boltzmann-Gibbs (BG) distribution would be the statistical base of the collision theory. In this way, the molecular velocity distribution is described by the Maxwellian exponential law,

$$f(u) = \left(\frac{m}{2\pi k_B T}\right)^{3/2} \exp\left(-\frac{mu^2}{2k_B T}\right), \tag{4}$$

where $m$ is mass of a molecule, and $u$ is velocity of a molecule.

Generally speaking, a chemical reaction is not in a thermodynamic equilibrium state but usually in a nonequilibrium state. In the reaction rate theory, what we are interested in is the processes of the evolution from one metastable state to another neighboring state, thus the assumption of thermodynamic equilibrium in the collision theory is quite farfetched. Very frequently, a system far away from equilibrium does not relax to a thermodynamic equilibrium state with a BG distribution, but might asymptotically approach a stationary nonequilibrium state with power-law distribution. In this situation, the Maxwellian distribution (4) should be replaced by the power-law one (2). In nonextensive statistical mechanics, the power-law distribution (2) can be derived using the extremization of Tsallis entropy [5]. In stochastic dynamical theory of Brownian motion in a complex system, Eq.(2) can also be obtained by solving the Fokker-Planck equations [4, 27]. When BG statistics is generalized to nonextensive statistics, the usual exponential and logarithm can be replaced by the $q$-exponential and the $q$-logarithm [5], respectively. Here the $\nu$-exponential [28, 29] can be defined as

$$\exp_\nu x = [1+(\nu-1)x]^{\frac{1}{\nu-1}}, \quad (\exp_1 x = e^x), \tag{5}$$

if $1+(\nu-1)x > 0$ and as $\exp_\nu x = 0$ otherwise. And the inverse function, the $\nu$-logarithm can be defined as

$$\ln_\nu x = \frac{x^{\nu-1}-1}{\nu-1}, \quad (x>0, \ln_1 = \ln x). \tag{6}$$

In this framework, along a Maxwellian path to the nonextensive velocity distribution [30], one can obtain the power-law velocity distributions. Thus the molecular velocity distribution Eq.(4) can be replaced by

$$f(u) = Z_\nu \left(\frac{m}{2\pi k_B T}\right)^{3/2} \left[1-(\nu-1)\frac{mu^2}{2k_B T}\right]^{\frac{1}{\nu-1}}, \tag{7}$$

where [6],

$$Z_\nu = \begin{cases} (\nu-1)^{5/2}\Gamma\left(\frac{1}{\nu-1}+\frac{5}{2}\right)\Big/\Gamma\left(\frac{1}{\nu-1}\right), & \nu>1, \\ (1-\nu)^{3/2}\Gamma\left(\frac{1}{1-\nu}\right)\Big/\Gamma\left(\frac{1}{1-\nu}-\frac{3}{2}\right), & \frac{1}{3}<\nu<1. \end{cases} \tag{8}$$

In the limit $\nu \to 1$, Eq.(7) is reduced to Eq.(4). Here, it would be helpful to introduce the physical meaning of the power-law parameter $\nu \ne 1$. In 2004, an equation of the parameter $\nu \ne 1$ was found both in the self-gravitating system and in the plasma system, and hence a clear physical explanation for $\nu \ne 1$ was presented [6, 7]. The equation can be written as $k_B \nabla T(r) = -(\nu-1)m\nabla\varphi_g(r)$ for the self-gravitating system [6] and $k_B \nabla T(r) = (\nu-1)e\nabla\varphi_C(r)$ for the plasma system [7], where $T(r)$ is space-dependent temperature, $m$ is mass of the particle, $e$ is charge of an electron, $\varphi_g(r)$ is the gravitational potential, and $\varphi_C(r)$ is Coulombian potential. The equation shows that the power-law distribution can be a nature of an interacting many-body system being at a nonequilibrium stationary-state. For a chemical reaction system, the equation of the $\nu$-parameter should be similar to that for the self-gravitating system, and $\varphi_g(r)$ might be considered as an intermolecular interaction potential.

Using the power-law velocity distribution function (7), the mean velocity of the molecules of a gas A is then given by

$$\bar{u} = 4\pi\int_0^\infty u^3 f(u)dv = 4\pi Z_\nu \int_0^\infty u^3 \left[1-(\nu-1)\frac{mu^2}{2k_B T}\right]^{\frac{1}{\nu-1}} du$$
$$= \chi_\nu \sqrt{\frac{8k_B T}{\pi m}}, \tag{9}$$

with the $\nu$–dependent parameter,

$$\chi_\nu = \begin{cases} (\nu-1)^{-\frac{1}{2}}\Gamma\left(\frac{1}{\nu-1}+\frac{5}{2}\right)\Big/\Gamma\left(\frac{1}{\nu-1}+3\right), & \nu>1, \\ (1-\nu)^{-\frac{1}{2}}\Gamma\left(\frac{1}{1-\nu}-2\right)\Big/\Gamma\left(\frac{1}{1-\nu}-\frac{3}{2}\right), & \frac{1}{2}<\nu<1. \end{cases}$$

Consequently, using $\bar{u}_{AB}^2 = \bar{u}_A^2 + \bar{u}_B^2$, the relative mean molecular velocity of molecules of types A and B is

$$\bar{u}_{AB} = \chi_\nu \sqrt{\frac{8k_B T}{\pi\mu}} \tag{10}$$

with the reduced mass $\mu$. Introducing (10) in Eq.(3) and approximately using $u_A = u_{AB}$, we obtain

$$Z_{AB} = \pi d_{AB}^2 \bar{u}_{AB} \frac{N_A N_B}{V^2}. \tag{11}$$

If we consider a gas that contains only molecules of type A, the total number of collisions would be

$$Z_{AA} = \pi d_{AA}^2 \bar{u}_{AA} \frac{N_A^2}{2V^2} = 2\pi r_A^2 \bar{u}_{AA} \frac{N_A^2}{V^2}, \tag{12}$$

where the factor 1/2 appears because we cannot count the same molecule twice. Since the relative mean velocity of the molecules of a gas is related to the mean velocity of the molecules, $\bar{u}_{AA} = \sqrt{2}\bar{u}_A$, Eq.(12) can be rewritten as

$$Z_{AA} = 2\sqrt{2}\pi r_A^2 \bar{u}_A \frac{N_A^2}{V^2}. \tag{13}$$

The expressions of collision density can be expressed in terms of pressures or molar concentrations, then the encounter rate coefficient for a bimolecular reaction between molecules of A and B is,

$$k_{AB} = \sigma_R \bar{u}_{AB} = \pi d_{AB}^2 \chi_\nu \sqrt{\frac{8k_BT}{\pi\mu}}. \tag{14}$$

Not every collision leads to reaction, so the encounter rate coefficient is always much bigger than the reaction rate coefficient. We need a certain amount of energy to convert reactants into products, and not all the collisions have enough energy to produce this chemical transformation.

Once more, we start from the power-law distribution Eq.(7) to calculate this factor. In practice, in collision theory we normally use the distribution function of molecular velocities in two dimensions rather than in three. Then the number $dN$ of the particles with the velocity ranging from $u$ to $u+du$ is

$$\frac{dN}{N_0} = \frac{\nu m}{k_BT}\left[1-(\nu-1)\frac{mu^2}{2k_BT}\right]^{\frac{1}{\nu-1}} u\,du, \tag{15}$$

where $N_0$ is the particle number density. We can rationalize this decrease in dimensionality by considering that at the moment of the collision between two molecules, the velocity vectors have a common point such that they lie within a plane. Thus, the velocity components within two dimensions that define this plane are sufficient to describe an effective collision. Based on the above equation, the number $N(\varepsilon)$ of the particles with the energy ranging from $\varepsilon$ to $\varepsilon+d\varepsilon$ is

$$\frac{dN(\varepsilon)}{N_0} = \frac{\nu}{k_BT}\left[1-(\nu-1)\frac{\varepsilon}{k_BT}\right]^{\frac{1}{\nu-1}} d\varepsilon. \tag{16}$$

By integration we can obtain the fraction of collisions for the energy equal to or more than the critical value $\varepsilon_c$ at which the reaction occurs,

$$\int_{\varepsilon>\varepsilon_c} \frac{dN(\varepsilon)}{N_0} = \left[1-(\nu-1)\frac{\varepsilon_c}{k_BT}\right]^{\frac{1}{\nu-1}+1}. \tag{17}$$

If this energetic term is included in the rate coefficient, the collision theory rate coefficient for the power-law distribution is derived by

$$k_\nu = \pi d_{AB}^2 \chi_\nu \sqrt{\frac{8k_B T}{\pi \mu}} \left[1 - (\nu-1)\frac{\varepsilon_c}{k_B T}\right]^{\frac{1}{\nu-1}+1}. \tag{18}$$

The Arrhenius pre-exponential factor can be defined as,

$$A_\nu = \pi d_{AB}^2 \chi_\nu \sqrt{\frac{8k_B T}{\pi \mu}}. \tag{19}$$

The activation energy expressed in molar terms is given by

$$E_a = -\frac{d\ln(k_\nu)}{d(1/RT)} = \frac{\nu E_c RT}{RT - (\nu-1)E_c} + \frac{RT}{2}. \tag{20}$$

As expected, when taking the limit $\nu \to 1$, they all become the familiar forms in the collision theory rate [1]:

$$k_{col} = \pi d_{AB}^2 \sqrt{\frac{8k_B T}{\pi \mu}} \exp(-\beta \varepsilon_c), \tag{21}$$

$$A = \pi d_{AB}^2 \sqrt{\frac{8k_B T}{\pi \mu}}, \tag{22}$$

and

$$E_a = -\frac{d\ln(k_{col})}{d(1/RT)} = E_c + \frac{RT}{2}, \tag{23}$$

### 3. Numerical analyses of the power-law collision theory rate coefficient

In order to illustrate dependence of the power-law collision theory rate coefficient on the physical quantities such as the parameter $\nu \neq 1$, the temperature $T$, and the critical energy $E_c$, we have made numerical analyses of $k_\nu$ with regard to $\nu$, $T$, and $E_c$, respectively, and have studied the variation of the generalized collision theory rate coefficient $k_\nu$ in Eq.(18) as a function of these quantities. In these numerical analyses, when one of these quantities was chosen as a variable, the other quantities were fixed. The fixed data were taken as typical data in chemical reactions. In this way, we have chosen $E_c = 20$ kJ mol$^{-1}$ and $T=300$K as the fixed values of the critical energy and the temperature, and have chosen $\mu = 10^{-26}$kg and $d_{AB}=10^{-10}$m as the fixed values of the reduced mass and the distance between A and B when a collision occurs.

Fig.1 has shown the dependence of the rate coefficient $k_\nu$ on the parameter $\nu$. The $k_\nu$-axis was plotted on a logarithmic scale. The range of the $\nu$-axis was chosen near 1.00, implying a state not very far away from the equilibrium. In Fig.1, the numerical analyses showed a very strong dependence of the generalized collision theory rate coefficient $k_\nu$ on the parameter $\nu$, which imply that a tiny deviation from the BG distribution and thus from the thermal equilibrium would result in a significant variation in the reaction rate. Such high sensitivity of the reaction rate to the $\nu$-parameter has shown the important role of the power-law distribution in the calculation of reaction rate coefficient, and again has told us that the nonequilibrium may be a key factor considered in the construction of the reaction rate theory.

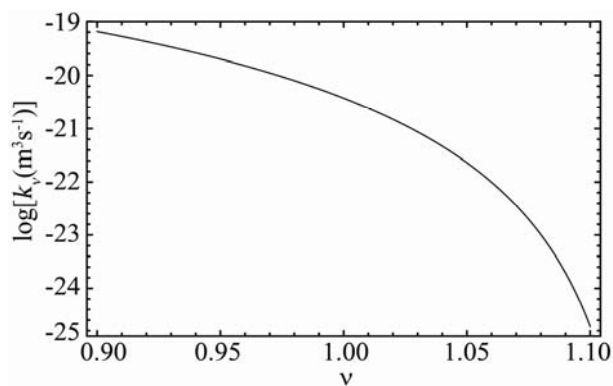

Fig.1. Dependence of the rate coefficient $k_v$ on the parameter $v$

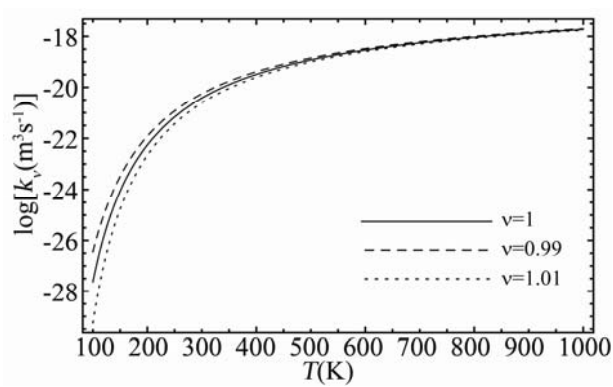

Fig.2. Dependence of the rate coefficient $k_v$ on temperature $T$ for three values of $v$

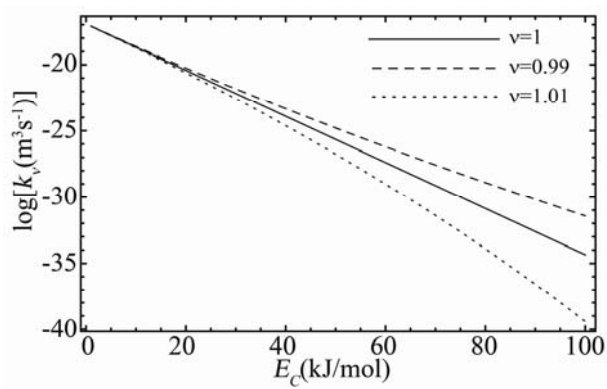

Fig.3. Dependence of the rate coefficient $k_v$ on $E_c$ for three values of $v$

    Fig.2 illustrated the dependence of the generalized collision theory rate coefficient $k_v$ on temperature $T$ for three values of $v$. The range of $T$-axis was chosen 100~1000K, as the typical temperature range in chemical reactions.

    Fig.3 illustrated the dependence of the generalized collision theory rate coefficient $k_v$ on the critical energy $E_c$ for three values of $v$. The range of $E_c$-axis was chosen 0~100kJ/mol, being in the order of the fixed value $E_c = 20$kJ/mol.

In the Figs.2 and 3, the curves of $v=1$ are corresponding to the conventional collision theory reaction rate coefficient in BG statistics.

## 4. Application to the mechanism of Lindemann-Christiansen

As an application, we analyze the unimolecular rate theory. In principle, unimolecular reaction is one of the most conceptually simple reactions in chemistry. However, the history of the understanding of simple decomposition processes has been one of the considerable debate, controversy and complexity even till our own days. One can imagine that even a true elementary reaction, such as the decomposition of $Br_2$, $Br_2 \to 2Br$, is impossible to have a simple collisional explanation although it shows a first-order rate law.

As a basis for all the modern theories of unimolecular reactions, the mechanism of Lindemann-Christiansen [26] has considered that the formation of a metastable molecule $A^*$ has sufficient energy to undergo reaction. The energization process involves collisions between two molecules of A [1],

$$A+A \underset{k_{-1}}{\overset{k_1}{\rightleftharpoons}} A^*+A, \quad A^* \xrightarrow{k_2} \text{products}. \tag{24}$$

The rate of product formation, i.e. the total reaction rate can be obtained by

$$v = \frac{d[P]}{dt} = k_2[A^*] = \frac{k_1 k_2 [A]^2}{k_2 + k_{-1}[A]}. \tag{25}$$

where [P] is the concentration of the products, [A] and [$A^*$] are the concentrations of A and $A^*$, respectively. In the vapour phase, at high pressures, one has $k_{-1}[A^*] \gg k_2$, and the rate of reaction assumes a more simple form,

$$v_\infty = \frac{k_1 k_2}{k_{-1}}[A] = k_\infty^1 [A], \tag{26}$$

where $k_\infty^1 = k_1 k_2 / k_{-1}$, which can be experimentally estimated at high pressures. The reaction is a first-order process.

A special concentration $[A]_{1/2}$ is defined at which the rates of both de-energization and product formation are equal [1],

$$k_{-1}[A^*][A]_{1/2} = k_2[A^*]. \tag{27}$$

However, the experimental values of $[A]_{1/2}$ are always much smaller than those predicted by the expression

$$[A]_{1/2} = \frac{k_2}{k_{-1}} = \frac{k_\infty^1}{k_1}, \tag{28}$$

when $k_1$ is estimated based on conventional collision theory rate. Thus, a modification must be contemplated to allow $k_1$ to be much larger and to account for the increase in that discrepancy with the increase in complexity of the reactants.

Using the new rate coefficient Eq.(18) to calculate $k_1$, we find that the value of $k_1$ can be significantly modified. As shown in Fig.1, the difficulty in Lindemann– Christiansen mechanism can overcome easily by using the power-law reaction rate coefficient for the reactions taking place in a nonequilibrium system with the power-law distributions. As we have pointed out, a tiny deviation of the $\nu$-parameter from 1 and thus from a BG distribution would result in significant changes in the rate coefficient. And the chemical reaction systems are of course away from equilibrium, otherwise there are no reactions at all, and the parameter $\nu \neq 1$ measures the distance away from equilibrium.

The second difficulty with the Lindemann–Christiansen mechanism is apparent when experimental data are plotted in another way [1],

$$\frac{1}{k^1} = \frac{k_{-1}}{k_1 k_2} + \frac{1}{k_1[\text{A}]}, \qquad (29)$$

where $k^1$ is a first-order rate coefficient, which is defined by $\upsilon = k^1[\text{A}]$. A plot of $1/k^1$ against the reciprocal of [A] should give a straight line. However, deviations from linearity of the kind have been found. Dealing with this difficulty with the Lindemann–Christiansen mechanism is a little complicated. First we rewrite Eq.(29) as,

$$\frac{1}{k^1} = \frac{1}{k_1}\left(\frac{k_{-1}}{k_2} + \frac{1}{[\text{A}]}\right). \qquad (30)$$

We can see from Eq.(24), $k_1$ is associated with the collision between two molecules A and A, $k_{-1}$ is associated with the collision between two molecules A and A$^*$, and $k_2$ has nothing to do with collision. It is a change of structure inside the molecule A$^*$.

As the pressure increases, the equilibrium involving A is more and more substantial, but the equilibrium involving A$^*$ is farfetched in some degree. This is because the reaction is consuming A$^*$, not A. The high pressure can ensure enough A to reach a state very closed to equilibrium; however, for A$^*$, as long as reactions occur, A$^*$ can not be in equilibrium, even far away from equilibrium. So when [A] increases, the equation of calculating $k_1$ is transitioning from Eq.(18) to Eq.(21), but the equation of calculating $k_2$ should always be Eq.(18). Then when the parameter $\nu$ is a little smaller than 1, the value from Eq.(18) are much more than that from Eq.(21), and thus $k_2$ is bigger, $k^1$ is smaller and the difficulty is overcome.

**5. Application to the pre-exponential factors**

In some chemical reactions, the experimental values have orders of magnitude smaller than or bigger than the calculated values [31]. It has been suggested that the geometric factor can be used to correct the deficiencies, but such corrections tend to be purely empirical, providing little physical insight [32]. In other words, the geometric factor is so small that it can not be understood as purely orientation effect. In fact, the geometric factor has lost its original meaning as a favorable orientation for reaction happens. Most geometrics are very difficult or even impossible to be calculated theoretically, even if they can, there are still differences between the experimental data and calculated values [32]. In order to overcome this flaw rooted in the collision theory and illustrate the generalized collision theory to a nonequilibrium system with the power-law $\nu$-distribution, we take three chemical reactions as examples to calculate the pre-exponential factors. The reason we calculate the pre-exponential factors instead of the rate coefficients is that the collision theory can not predict the activation energy [33].

The pre-exponential factors of the reactions in the experimental studies were taken from the NIST chemical kinetics database at http://kinetics.nist.gov/kinetics. The activation energies are taken from recent reviews of the relative reactions. In Table 1, we listed the pre-exponential factors of three reactions, where $A_{\text{exp}}$ is the experimental values, $A_{\text{col}}$ is the old theoretical values calculated using Eq.(22) (the detailed calculations can be found in [34]) and $A_\nu$ is the new theoretical values calculated using Eq.(19). The activation energies $E_a$ and the pre-exponential factors, $A_{exp}$, $A_{col}$ and $A_\nu$, are given in unit of kJmol$^{-1}$ and dm$^3$mol$^{-1}$s$^{-1}$ respectively. All of the values in this table are given at the temperature of $T$=300K. The quantity $\delta$ is a relative error of $A_{col}$ to $A_{exp}$, defined by $\delta = |(A_{col} - A_{exp})/A_{exp}|$.

Table 1. Experimental and theoretical values of the pre-exponential factors

| Reaction | $E_a$ | $A_{col}$ | $A_{exp}$ | $\delta$ | $A_\nu$ | $\nu$ |
|---|---|---|---|---|---|---|
| F+H$_2$→FH+H | 3.74 | 1.5×10$^{11}$ | 7.8×10$^{10}$ | 92% | 7.8×10$^{10}$ | 1.16 |
| CO+O$_2$→CO$_2$+O | 264 | 1.4×10$^{11}$ | 2.1×10$^8$ | 66567% | 2.1×10$^8$ | 1.001 |
| CH$_3$+CH$_3$→C$_2$H$_6$ | 11.56 | 2.5×10$^{11}$ | 1.1×10$^{10}$ | 2173% | 1.1×10$^{10}$ | 1.12 |

Table 1 shows very significant relative errors of the old theoretical values $A_{col}$ to the experimental values $A_{exp}$. For these three chemical reactions, however, we find that the new theoretical values $A_\nu$ with the corresponding $\nu$-parameter slightly different from 1 can be in exactly agreement with the values in all the experimental studies. According to the physical meaning of the $\nu$-parameter deviating from 1[6,7], it is shown that the new Arrhenius pre-exponential factor Eq.(19) in the collision theory rate coefficient for the power-law distribution may be one of important candidate more suitable for describing the nonequilibrium chemical reaction rates.

**6. Conclusions and discussions**

The reaction rate theory for the systems with power-law distributions is beyond the scope of conventional collision theory for the systems with a BG distribution, and therefore if chemical reactions occur in the systems with power-law distributions the collision theory reaction rate formulae need to be modified. The purpose of this work is to generalize the conventional collision theory rate formula Eq.(21) to a nonequilibrium system with the power-law $\nu$-distribution. The present work goes only along a classical statistical theory, and the approach to generalize the collision theory rate formula follows the standard line in textbooks.

In conclusion, we have studied the collision theory reaction rate coefficient for the power-law $\nu$-distribution. We have derived a generalized collision theory reaction rate formula Eq.(18) to the reactions taking place in a nonequilibrium system with the power-law $\nu$-distribution, which as compared with the old formula Eq.(21) depends not only on the temperature $T$ and critical energy $E_c$, but also on the power-law $\nu$–parameter.

We have made numerical analyses to illustrate the dependence of the power-law collision theory rate coefficient $k_\nu$ on the relevant physical quantities. We clearly showed a very strong dependence of $k_\nu$ on the power-law $\nu$-parameter different from 1, and indicated that a tiny deviation from BG distribution would result in a very significant effect on the reaction rate coefficient. Such high sensitivity of the reaction rate coefficient to the $\nu$–parameter showed that the power-law distributions play an important role in the calculations of reaction rate coefficients, and that the nonequilibrium is a key factor to be considered in the construction of the reaction rate theory for a complex system.

We have applied the power-law collision theory to the unimolecular rate theory to make an explanation that it can successfully overcome the two difficulties of the Lindemann–Christiansen mechanism, which is the basis for all the modern theories of unimolecular reactions.

We also applied the new Arrhenius pre-exponential factor Eq.(19) appearing in the power-law collision theory rate coefficient Eq.(18) to calculate the pre-exponential factors for three examples of the reactions taking place in nonequilibrium systems with the power-law $\nu$-distributions. As compared with the old theoretical formula, we showed that the new pre-exponential factor Eq.(19) with the $\nu$-parameter slightly different from 1 can be in exactly agreement with the values measured in the experimental studies.

**Acknowledgment**

This work is supported by the National Natural Science Foundation of China under Grant No. 11175128 and also by the Higher School Specialized Research Fund for Doctoral Program under Grant No. 20110032110058.